\documentclass[aps,prl,reprint,superscriptaddress,longbibliography]{revtex4-1}
\usepackage{amsmath}
\usepackage{amsfonts}
\usepackage{color}
\newcommand{ \ba}{\begin{eqnarray}}
\newcommand{ \ea}{\end{eqnarray}}

\usepackage{epsf,latexsym,graphicx}
\usepackage{xr}
\usepackage{xcite}
\externalcitedocument{NFLHolstein_supp}

\begin{document}

\title{ Non-Fermi liquid over extended range at zero temperature without quantum criticality }

\author{Tae-Ho Park}
\affiliation{Department of Physics and Institute for Basic Science Research, Sungkyunkwan University, Suwon 16419, Korea}

\author{Han-Yong Choi}
\affiliation{Department of Physics and Institute for Basic Science Research, Sungkyunkwan University, Suwon 16419, Korea}

\date{\today}

\begin{abstract}

A strange metal state appears in many strongly correlated materials from the cuprates, pnictides, and twisted bilayer graphenes to even bosonic systems. Enormous efforts are being made to unravel the nature of the non-Fermi liquid (NFL) which underlies the strange metallicity, yet progress is rather slow. Understanding the NFL behavior is crucial, in addition to its importance in replacing the Fermi liquid paradigm, in that NFL is the normal state from which new states of matter emerge like high temperature superconductivity. 
Here, we report the appearance of an NFL phase in a discontinuous metal-NFL-insulator transition in a ``non-engineered'' Hamiltonian of the Holstein model away from half-filling.
The discontinuity of the phase transition between NFL and Fermi liquid metal renders the ground state at the phase boundary a mixture between them over an extended range like the extended criticality observed in cuprates and other quantum materials. The dynamical mean-field theory was employed in combination with Wilson's numerical renormalization group technique. We identify the origin of the NFL metal phase as the opening of a spin gap in the absence of a charge gap. Also, strong lattice fluctuations emerge at the boundary which should enhance superconductivity.

\end{abstract}


\maketitle

Right after the discovery of high temperature superconductivity in cuprates, it was realized that they exhibit, perhaps more surprisingly, strange metal (SM) behaviors in the normal state. This is probably most apparent in the way the electrical resistivity $\rho$ changes with the temperature $T$ where $\rho$ is linear in $T$ down to the low $T$ limit and up to high $T$ melting temperature\cite{Varma_RMP2020,Proust_ARCP2019} in contrast to $\rho \sim T^2$ for the Fermi liquid metal (FLM). Despite the enormous efforts, progress is rather slow, and developing a theoretical framework to understand the NFL physics thus far remains one of central problems in modern condensed matter physics. There are several theoretical frameworks that lead to NFL behavior:\cite{Phillips_Science2022,Chowdhury_RMP2022,SSLee_ARCMP2018} (a) It can arise when electrons are coupled to quantum critical bosonic fluctuations, which often goes under the name of Hertz-Millis-Moriya criticality. (b) A different form of quantum criticality is associated with a change of the Fermi surface volume. This includes continuous metal-insulator transitions and Kondo-breakdown transitions. (c) Sachdev-Ye-Kitaev-like models where fermions interact through random and all-to-all couplings.\cite{Chowdhury_RMP2022} In these frameworks, the SM phase resides at a single quantum critical point of a continuous quantum phase transition at zero temperature, and fans out as the temperature increases.

Experimentally, on the other hand, it was found in many quantum materials that the $T$-linear resistivity remains robust down to the low $T$ limit over a finite range of external parameters like doping.\cite{Cooper_Science2009,Phillips_Science2022,Jiang_NaturePhys2023} Such SM of an extended criticality was further identified by the observation of the $H$-linear magnetoresistance at different doping levels of overdoped cuprates.\cite{Ayres_Nature2021} In manganese silicide, Pfleiderer $et~al$. reported that a stable extended NFL state emerges under applied pressure without quantum criticality.\cite{Pfleiderer_Science2007} Hu $et~al$ observed that the NFL behavior remains robust after the full suppression of the field-induced quantum critical point in CeCoIn$_5$ by Yb doping.\cite{Hu_PNAS2013} These observations suggest that the NFL behavior could be a new state of matter in its own right rather than a consequence of the underlying quantum criticality.

In contrast, controlled microscopic theories to produce SM over an extended region are scarce, and only recently do they begin to be reported. Wu $et~al$ investigated the two-dimensional Hubbard model using the dynamical mean-field theory (DMFT).\cite{Wu_PNAS2022} They found that the SM phase of linear $T$ resistivity resides in a finite range of doping down to low $T$ in between the pseudogap and Fermi liquid regimes. NFL behavior is due ultimately to an anomalous spectrum of gapless excitations which invalidates the phase-space arguments that protects the Fermi liquids. Recent works report that the SM behavior can be obtained over an extended range in a model of $N$ fermion species coupled to $M \sim N$ dynamical two-level systems per site via spatially random interactions in the large $N$ limit.\cite{Bashan_PRL2024,Tulipman_arx2024}
Despite these exciting developments, analytical or numerically exact solutions of ``natural/non-engineered'' microscopic models which establish the NFL as a ground state are of fundamental importance in understanding the NFL and emerging superconductivity.

Here, we report a new type of discontinuous metal-insulator transition where NFL intervenes in between them at zero temperature in the Holstein model away from half-filling.
The discontinuity of the phase transition between the metal and NFL renders the ground state a mixture between them and the strange metallicity of a two-fluid kind appears over an extended range at the phase boundary. Amusingly, this is not unlike the two-fluid model envisaged for the extended criticality observed in cuprates \cite{Cooper_Science2009,Ayres_Nature2021,Ayres_FrontPhys2022,Hussey_PhysicaC2023,Bozovic_Nature2016,Bozovic_PhysicaC2019}, twisted bilayer graphenes \cite{Jaoui_NaturePhys2022}, and twisted transition metal dichalchogenides \cite{Ghiotto_Nature2021}.
The Holstein model is an archetype Hamiltonian of electron-phonon coupled systems as given by Eq.\ (\ref{eq:Hol}) below. We identify the origin of the NFL as the opening of the spin gap of onsite spin singlet pairs (bipolaron) in the absence of the charge gap. Being onsite bipolarons, the NFL here is a bosonic metal, or it is a spin gap metal (SGM).
The insulating state is the onsite bipolaron insulator as has been discussed before \cite{jeon_2004}, and is termed spin gap insulator (SGI) here to be contrasted with the SGM.



The NFL phase was identified from calculations of the self-energy $\Sigma(\omega)$. DMFT was employed in combination with Wilson's numerical renormalization group (NRG) technique to solve the Holstein model in infinite dimensions. The quasiparticle weight $z$ is defined by
 \begin{equation}\label{eq:z}
z = \left[ 1- \frac{\partial}{\partial \omega} Re \Sigma(\omega) \right]_{\omega \rightarrow 0}^{-1} ,
 \end{equation}
where $Re$ means the real part. $z$ has a physical meaning of quasiparticle weight only for $0 < z \leq 1$, for which a phase is a Fermi liquid metal (FLM). Otherwise, it is NFL metal if the renormalized single particle density of states (DOS) at the Fermi energy $\rho(\omega=0) \ne 0$, or insulator if $\rho(0) = 0 $. Furthermore, by calculating the spin-spin correlation function and spin gap $\Delta_\sigma$, the NFL is identified as a spin gap metal phase. See Fig.\ 4 below.

The Holstein model is given by
\begin{eqnarray} \label{eq:Hol}
{\cal H} = -t  \sum_{\langle i,j\rangle\sigma}
c_{i\sigma}^\dag c_{j\sigma} -\mu \sum_i {\hat n}_i +\omega_0 \sum_i a_i^\dag a_i
 \nonumber \\
+ g \sum_i ( a_i^\dag +a_i) \left( {\hat n}_i -1 \right),
\end{eqnarray}
where $c_{i\sigma}$ and $a_i $ are the fermion of spin $\sigma$ and boson operators at the site $i$, and $\langle i,j\rangle$ implies the nearest neighbors, ${\hat n}_i =\sum_\sigma c_{i\sigma}^\dagger c_{i\sigma}$ the electron density operator at the site $i$, and $\mu$ is the chemical potential to control the electron density of the conduction band $n$. $\mu=0$ corresponds to the half-filling ($n=1$) and $\mu<0$ covers the less than half-filling ($n<1$).
The electrons are coupled with the Einstein phonon of frequency $\omega_0$ with the onsite coupling constant $g$. The model has been explored previously via various tools including the DMFT\cite{benedetti_1998,meyer_2002,capone_2003,jeon_2004,koller_2004,georges_1996}.
It has been a challenge to treat theoretically the different energy scales of electrons and phonons and emergent small energy scales of the metal-insulator transition, superconductivity, and soft boson modes. The NRG can zoom in, via a logarithmic discretization of the conduction band, onto the low lying states with an arbitrary precision and is ideal to address the small energy scales in this problem. We take the half bandwidth $D$ as the unit of energy which equals the Fermi energy $\epsilon_F$ at the half-filling.

\begin{figure}[h]
\begin{center}
\includegraphics[width=\linewidth]{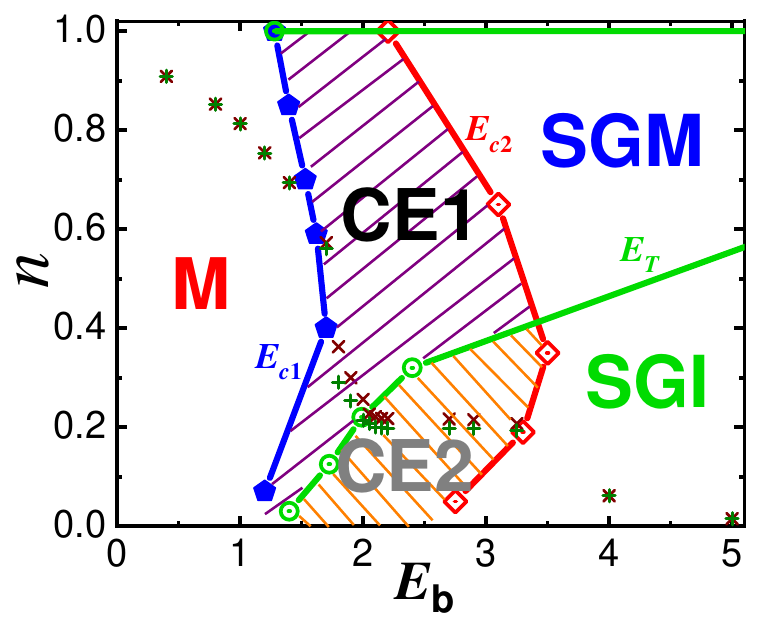}
 \caption{(a) The schematic normal state phase diagram of the Holstein model at zero temperature in the plane of $E_b$ and $n$ for $\omega_0 = 2 D$. There appear the Fermi liquid metal (FLM), spin gap metal (SGM) and spin gap insulating (SGI) phases. The phase boundaries between them are represented by the blue pentagons ($E_{c1}$), red rhombuses ($E_{c2}$), and green circles ($E_T$). The lines connecting them are schematic and guide to eyes. The hatched area CE1 (CE2) permits two solutions of the FLM \& SGM (SGM \& SGI) of which SGM (SGI) is the ground state. $E_T$ is the metal-insulator (SGM-SGI) transition line. The green ($+$) and purple ($\times$) symbols are the electron densities $n$ calculated from the obtained solutions as $E_b$ is varied for the $\mu =- 0.03$ cut. See the main text for more details. }
\label{fig:phase}
\end{center}
\end{figure}

We first present a schematic phase diagram of the Holstein model of the normal phases with no broken symmetry at the zero temperature in the plane of the band filling $ n$ and the bipolaron energy $E_b = 2g^2/\omega_0$ (See the Method section below.) as shown in Fig.\ 1. Self-consistent NRG-DMFT calculations were done in the grand canonical ensemble, that is, the density $n$ is not fixed, but we fix the chemical potential $\mu$ and vary $E_b$, or fix $E_b$ and vary $\mu$.
The phase boundary lines $E_{c1}$ and $E_{c2}$ were determined according to whether there exist one or two solutions in the region just like the Mott transition in the repulsive Hubbard model.\cite{georges_1996} Two solutions exist for $E_{c1} <E_b <E_{c2}$ as shown as the hatched areas CE1 and CE2 in Fig.\ 1, and only a single solution otherwise.
A good way to obtain these two solutions is to start numerical iterations from metallic or insulating initial configurations for chosen $\mu$ and $E_b$.

\begin{figure*}[t]
\begin{center}
\includegraphics[width=0.85\linewidth]{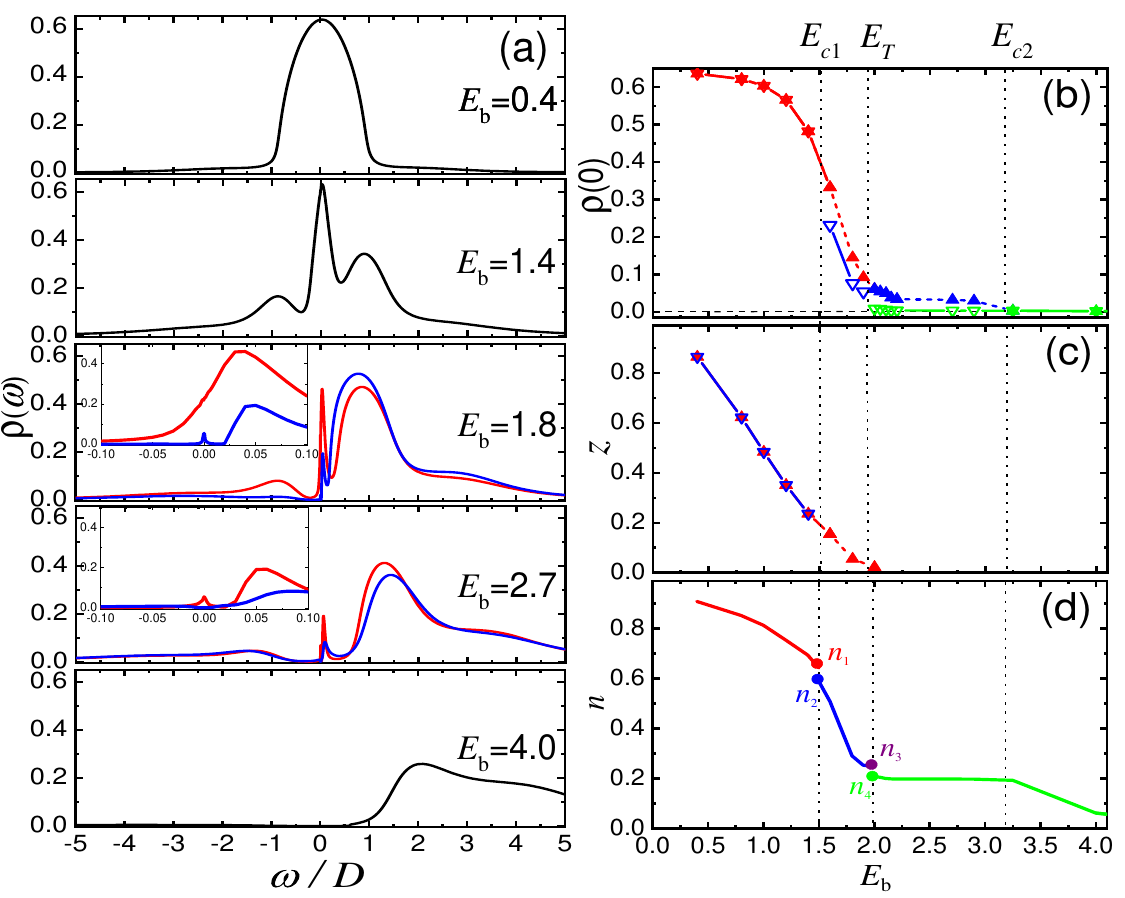}
 \caption{ (a) The renormalized DOS $\rho(\omega)$ for selected $E_b$ along the $\mu= -0.03$ cut. The red and blue curves represent the solutions obtained from metallic and insulating initial configurations which are, respectively, FLM \& SGM in CE1 and SGM \& SGI in CE2 region.  The plots, from top to bottom, belong to the M, M, CE1, CE2, and SGI regions in the phase diagram of Fig.\ 1.
 $\rho(0)$, $z$, and $n$ along the same cut are shown in (b), (c), and (d), respectively. The red, blue, and green data points in (b)--(d) represent results from the FLM, NFL, SGI. The electron density $n$ in Fig.\ 2(d) is shown for the ground state solutions, which corresponds to the green (+) symbols in the phase diagram of Fig.\ \ref{fig:phase}.
}
\label{fig:eDOS}
\end{center}
\end{figure*}

Following previous works \cite{hohenadler_2013,Balents_PRB1996}, we may classify the obtained phases with no broken symmetry in each region by the absence or presence of gaps for charge, spin, and single-particle excitations, ${\Delta}_\rho$, ${\Delta}_\sigma$, ${\Delta}_{sp}$, respectively.
Unlike problems in one-dimension, however, the boundaries of vanishing $\Delta_{sp}$ and $\Delta_\rho$ are the same in infinite-dimensions (see below after Eq.\ (\ref{spinspectrum})), hence, we will not distinguish between them here. Then, it is FLM for ${\Delta}_\rho = {\Delta}_\sigma = 0$, SGM for ${\Delta}_\rho = 0$ \& ${\Delta}_\sigma \ne 0$, and SGI for ${\Delta}_\rho \ne 0$ \& ${\Delta}_\sigma \ne 0$. The Mott insulator of ${\Delta}_\rho \ne 0$ \& $\Delta_\sigma = 0$ appears for the repulsive Hubbard model and is irrelevant here.\cite{hohenadler_2013}
Consider $\mu= -0.03 D$ as a representative cut as indicated by the green ($+$) and purple ($\times$) symbols in Fig.\ 1. Along the cut, we plot the evolution of $\rho(\omega)$, $\rho(0 )$, and $z$ in Fig.\ 2(a), 2(b), and 2(c), respectively.
As can be read off from the Fig.\ 2(a) or 2(b), both of the obtained solutions are metallic for $E_{c1} <E_b < E_T$ of the CE1 region, and one is metallic and the other is insulating for $E_{T} <E_b < E_{c2}$ of the CE2 region of Fig.\ 1. $E_T$ is the metal-insulator (SGM-SGI) transition line.
We determined that $E_{c1} \approx 1.6$, $E_{T} \approx 2.0$, and $E_{c2} \approx 3.2$ for $\mu=-0.03$.
The calculated $z$ from Eq.\ (\ref{eq:z}) falls within $0 < z \leq 1 $ only for the FLM phase as shown in Fig.\ 2(c). The other metallic solution in CE1 has the unphysical $z<0$ if calculated using Eq.\ (\ref{eq:z}), hence, is an NFL metal, as is discussed in more detail below in terms of the self-energy. In CE2, two solutions are NFL metal and SGI.

In the CE1 region, the NFL solution is a lower energy state than FLM and thus the ground state, and in CE2, the SGI solution is the ground state, as is shown in Fig.\ \ref{fig:energy} in the Supplementary Information (SI) \cite{supplement}.
Then, there occur two phase transitions between FLM \& NFL at $E_{c1}$ and between NFL \& SGI at $E_T$. Both are first order transitions where $\rho(0)$, $z$, and $n$ change discontinuously as can be seen, respectively, from Fig.\ 2(b), 2(c), and 2(d). More explicitly, the electron density $n$ decreases continuously as $E_b$ increases, and at $E_b =E_{c1}$ it jumps between $n_1$ and $n_2$. And, as $E_b$ further increases beyond $E_{c1}$, $n$ decreases below $n_2$, and at $E_b =E_T$ it jumps between $n_3$ and $n_4$.

\begin{figure}[h]
 \begin{center}
\includegraphics[width=0.85\linewidth]{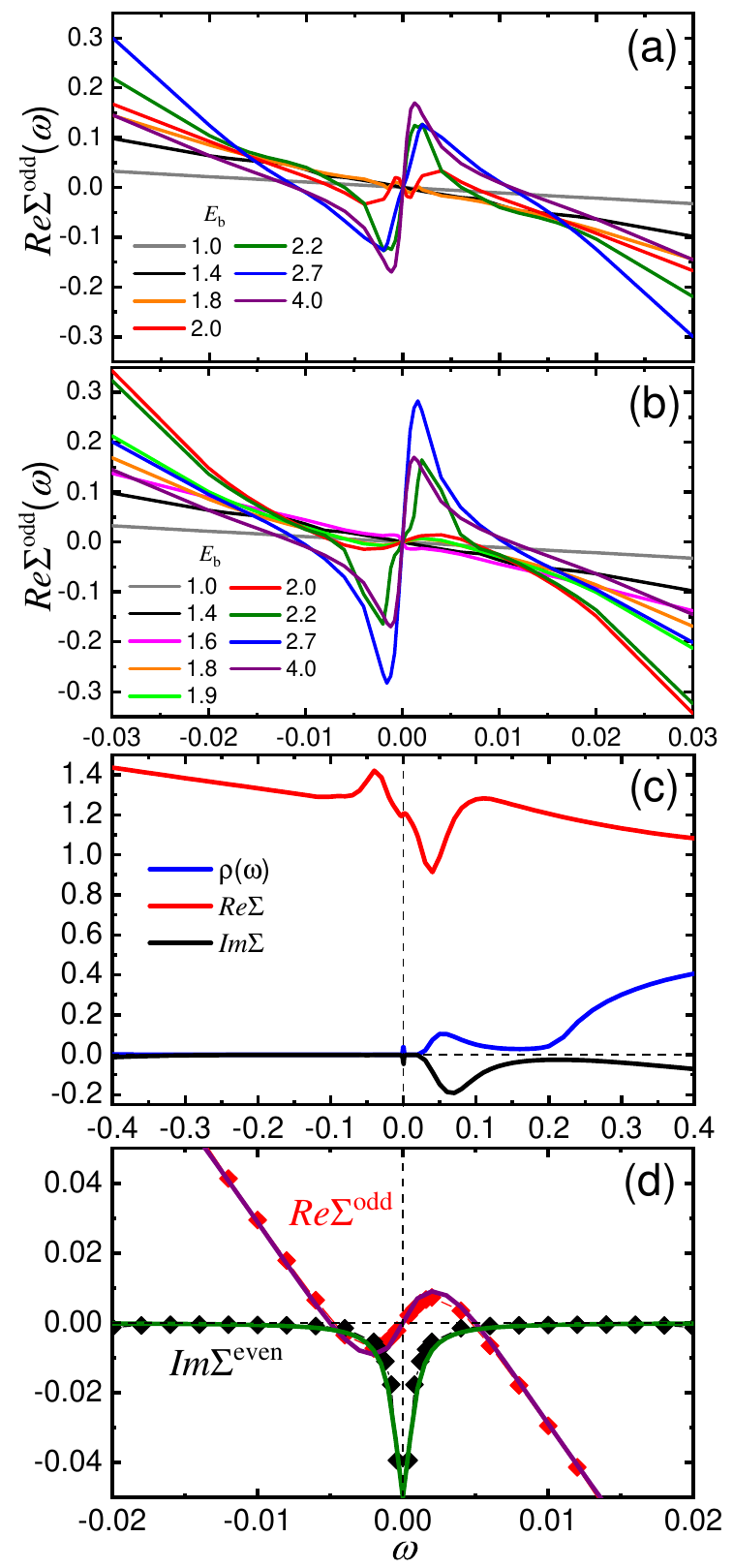}
 \caption{ (a) and (b) shows the results obtained via numerical iterations starting from the metallic and insulating initial configurations, respectively. $ \left. \partial Re\Sigma(\omega) /\partial\omega \right]_{\omega\rightarrow 0} >0 $ is the criterion of NFL.
 (c) The self-energy and DOS for SGM for $E_b = 1.9$. (d) is a zoom in on the $\omega \approx 0$ region of the symmetrized self-energy. The symbols are the numerical calculation results and the solid lines are the fitting curve of Eq.\ (\ref{NFL_self}). }
\label{fig:selfenergy}
 \end{center}
\end{figure}

As can be seen from Fig.\ 2(d), therefore, for the electron density $n_1 >n> n_2$ ($E_b =E_{c1}$), the ground state is a mixture of FLM and SGM. The SGM being onisite pairs (bipolarons), it has a bosonic character, and the ground state is akin to a fermion–boson mixture, where fermionic and bosonic excitations coexist in the normal state \cite{Bozovic_Nature2016,Mamood_PRL2019,Bozovic_PhysicaC2019,Bozovic_Quantum2018}. Also, this is not unlike the two-fluid model for the SM observed in the overdoped cuprates over the extended doping range between $p^* < p < p_{end}$, where $p^*$ and $p_{end}$ are, respectively, the doping concentrations at the pseudogap quantum critical point and at the end of the superconducting dome \cite{Cooper_Science2009,Ayres_Nature2021,Ayres_FrontPhys2022,Hussey_PhysicaC2023}. The resistivity measurements on cuprates for $p^* < p < p_{end}$ revealed that they can be best fit by
 \begin{eqnarray}
 \rho(T) -\rho_0 = \alpha_1 T + \alpha_2 T^2,
 \end{eqnarray}
where the SM coefficient $ \alpha_1 \rightarrow 0$ as $p \rightarrow p_{end}$ and $\alpha_2$ is the FLM contribution. This
dichotomy hints strongly at the presence of two distinct contributions
to the in-plane transport in overdoped cuprates, one described by
conventional transport theory, the other anomalous and characteristic of Planckian dissipation physics.

We now look at the emergence of the NFL phase in terms of the self-energy $\Sigma(\omega)$.
For a FLM, the slope of $Re \Sigma(\omega)$ around $\omega=0$ must be negative so that $0<z \leq 1$ from Eq.\ (\ref{eq:z}). To show this more clearly, we plot the symmetrized self-energy in Fig.\ \ref{fig:selfenergy}(a) and (b),
$\Sigma^{odd} (\omega) = \left[ \Sigma(\omega)-\Sigma(-\omega) \right]/2$ and $\Sigma^{even} (\omega) = \left[ \Sigma(\omega)+\Sigma(-\omega) \right]/2$.
As $E_b$ is increased, the negative slope becomes steeper and changes to a positive slope around $\omega \approx 0$ which is a well-known signal of NFL or non-metal.\cite{Bulla2001prb}
Fig.\ \ref{fig:selfenergy}(c) shows the real and imaginary parts of $\Sigma(\omega)$ for $E_b = 1.9$ in the CE1 region, and Fig.\ 3(d) is zoom in on the $\omega=0$ region. It demonstrates that the self-energy of NFL around $\omega \approx 0$ is very well represented by
 \begin{eqnarray}
 \Sigma_{NFL}(\omega) = \frac{A}{\omega^\alpha + i \delta^\alpha} ,
 \label{NFL_self}
 \end{eqnarray}
where $A$ is a constant, and $\alpha \approx 0.85$ and $\delta\approx 0.003 D$. The exponent $\alpha<1$ indicates the strange metallicity as is discussed below in Eq.\ (\ref{eq:resistivity}).
Interestingly, Xu $et~al$ reported from Quantum Monte Carlo calculations of a model describing Ising ferromagnetic fluctuations coupled to a Fermi surface that the leading term of the NFL self-energy due to thermal fluctuations is $1/\omega$ at a quantum critical point\cite{Xu_jpjQM2020}, while we have the purely quantum NFL self-energy of the form of Eq.\ (\ref{NFL_self}).

\begin{figure}[t]
\begin{center}
\includegraphics[width=0.85\linewidth]{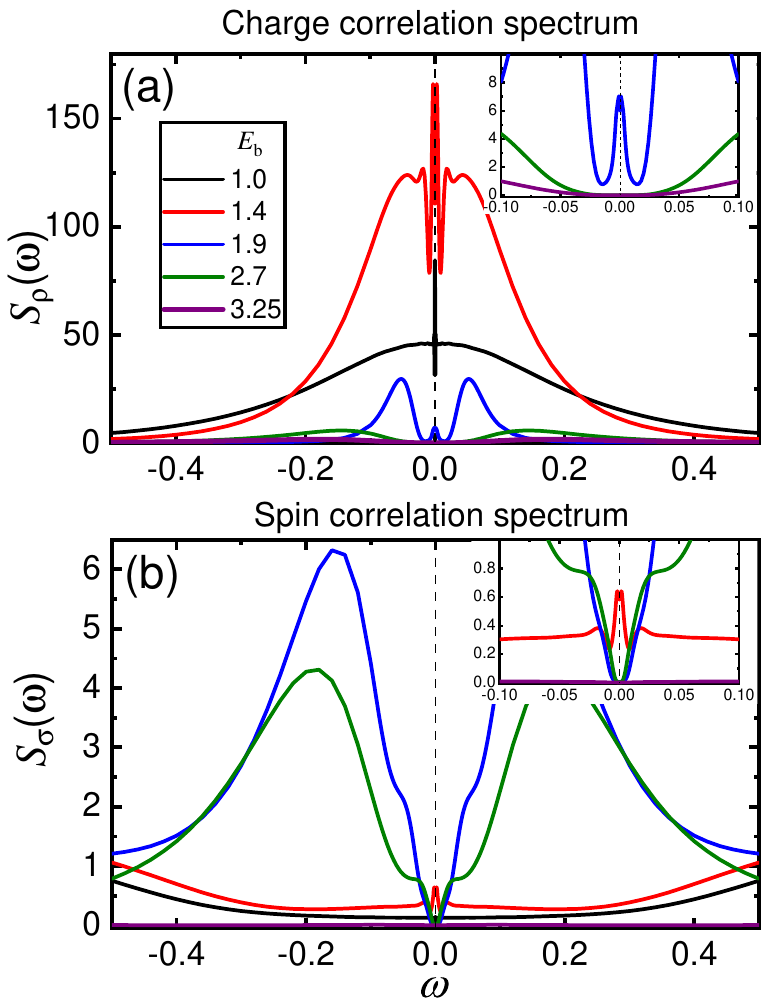}
 \caption{
 The charge and spin spectra for selected $E_b$ along the $\mu= -0.03$ cut calculated from the ground state solutions are shown in (a) and (b), respectively.  The insets are the zoom in on the low energy region to exhibit the gap feature more clearly. For $E_b = 1.9$ (the curves in blue) in the CE1 region, the charge spectrum shows no gap $\Delta_\rho =0$, while the spin spectrum shows the full gap of $\Delta_\sigma \approx 0.03$. This clearly establishes that the NFL is a spin gap metal.
}
\label{fig:spingap}
\end{center}
\end{figure}

Now, we turn to the nature of the obtained NFL phase of the Holstein model.
Previous DMFT calculations combined with the quantum Monte Carlo or exact diagonalization for the attractive Hubbard model, which is closely related to the Holstein model, reported discontinuous pairing transitions between the FLM and spin-gapped phases.\cite{Keller_PRL2001,Capone_PRL2002} The energy resolution of the calculations was not sufficient to distinguish between metal and insulator. We therefore checked if the NFL metal in our calculations is indeed a spin gapped phase. We calculated the spin susceptibility as detailed in the section S3 in SI. The imaginary part determines the spin spectrum.
 \begin{eqnarray}
 S_\sigma (\omega) = -\frac{\pi}{\omega} Im \chi_\sigma (\omega).
 \label{spinspectrum}
 \end{eqnarray}
We show the spin spectrum $S_\sigma (\omega)$ in Fig.\ \ref{fig:spingap} calculated from the ground state solutions. It exhibits the spin gap $\Delta_\sigma \ne 0$  ($S_\sigma(0) = 0 $) in the SGI ($E_b =2.7, 3.25$) and gapless $\Delta_\sigma(0) = 0 $ in FLM ($E_b = 1.0, 1.4$) regions just like the charge spectrum $S_\rho (\omega)$ behavior. On the other hand, the NFL phase in the CE1 ($E_b = 1.9$) exhibits a spin gap $\Delta_\sigma \ne 0$, but a zero charge gap $\Delta_\rho = 0$, as can be read from Fig.\ \ref{fig:spingap}. Note that the existence and absence of the charge gap is the same as the single particle gap as can be checked from Fig.\ \ref{fig:eDOS}. This establishes that the NFL is indeed a spin gap metal phase. Two electrons form an onsite spin singlet pair ($i.e.$, onsite bipolaron), and the binding energy of the singlet is the spin gap $\Delta_\sigma$. Yet the charge channel is gapless because of the single particle hopping to lower the kinetic energy. This conductive spin gapped normal state was reported previously from the variational Monte Carlo study of the two-dimensional attractive Hubbard model.\cite{Tamura_JPSJ2012} But no discussion is given on the non-Fermi liquid nor insulating phase. The present work is on the Holstein model with the phonon frequency $\omega_0 =2D$ which is rather close to the attractive Hubbard model (See the Method section below.) It remains to be established if all three phases appear in the attractive Hubbard model as well like Fig.\ 1.

We note that the emergence of the NFL phase away from half-filling may be discussed in terms of the NRG valence fluctuation fixed point, thereby cementing the existence of NFL phase in addition to the metal and insulating phases. For technical details, see the comments \cite{fixed-point} and the NRG energy flow plots in Fig.\ \ref{fig:NRGflow} in SI.

It should, of course, be interesting to investigate what kind of strange metallicity the NFL of spin gap metal is to exhibit. For that the current $T=0$ calculations need to be extended to the finite temperature.
We may proceed to consider the transport properties using the Kubo formula. The $T$ dependence of the resistivity follows from the frequency dependence of the self-energy.
The dc conductivity is given by \cite{Cha_PNAS2020}
 \begin{eqnarray}
\sigma(T) \propto \frac{1}{T} \int d\omega \frac{1}{Im \Sigma(\omega)} {\rm sech^2} \left(\frac{\omega}{2T} \right) .
 \end{eqnarray}
We find using the self-energy of Eq.\ (\ref{NFL_self})
 \begin{eqnarray}
 \rho(T \ll \delta) - \rho_0 \propto T^\gamma ,
 \label{eq:resistivity}
 \end{eqnarray}
where $\gamma = 2\alpha \approx 1.7$, $\alpha$ being the exponent of the self-energy as discussed in the recent report of extended region of NFL \cite{Bashan_PRL2024,Tulipman_arx2024}

Interestingly, enhanced lattice fluctuations appear at $E_{c1}$, the phase boundary between the Fermi liquid metal and NFL, which should be favorable for superconductivity. This is presented in Fig.\ \ref{fig:latfluctuations} in the SI. The lattice fluctuations $ \left< x^2 \right> - \left< x \right>^2 $, where $ x \sim a+ a^\dag$ is the displacement operator, are proportional to the area under the Eliashberg function $\alpha^2 F(\omega)$, where $F(\omega)=\rho_{ph}(\omega)$ is the imaginary part of the phonon propagator discussed in S2 in the SI.
They become maximal accompanying the emergent soft phonon mode near the lower critical value of the metal-insulator transition as reported by Meyer {\it et al.} for the half-filled Holstein model.\cite{meyer_2002} We expect that the superconducting $T_c$ becomes maximal at $E_{c1}$ which is consistent with the report that $T_c$ is maximum at $U_{c1}$ for the attractive Hubbard model.\cite{toschi_2005b,toschi_2005}


We used $\left. \partial Re \Sigma(\omega)/\partial\omega \right]_{\omega \rightarrow 0} >0 $ at zero temperature as a criterion for non-Fermi liquid in this paper and reported the appearance of NFL which intervenes in the metal-insulator transition in the electron-phonon coupled Holstein model. In combination with the dynamical mean-field theory, the key features of the NRG technique of the non-perturbative, zero temperature, and arbitrary high precision were crucial in establishing that the NFL is a spin gap metal phase and that it is a ground state of the non-engineered Holstein model rather than a consequence of the underlying quantum criticality. The 1st order nature of the phase transition between the NFL and metal phase means that the ground state is a mixture between them over extended parameter range at the phase boundary. This is in line with the two-fluid model envisaged before to understand the strange metallicity observed in the cuprates in the overdoped region. Yet, it remains to be established through further theoretical and experimental studies if the discontinuity in the phase transition between metal and NFL indeed underlies the two-fluid like behavior observed in the overdoped cuprates and other quantum materials.

While most experimental observations of NFL behavior are on the strongly correlated electron systems, this should not invite an interpretation that the current work is irrelevant for real materials. It perhaps implies that the NFL is rather ubiquitous as suggested by the observation of strange metallicity in bosonic systems \cite{Yang_Nature2022} and the appearance of quantum spin liquid from electron-phonon coupled Su-Schrieffer-Heeger model \cite{Cai_arXiv2024}.
To the best of our knowledge, this work is the first report of NFL in a natural/non-engineered model, at least for electron-phonon coupling dominant model. We believe that this will provide a significant way to explore the long standing problem of how the NFL and high $T_c$ superconductivity are intertwined in quantum materials.

\begin{acknowledgments}

We acknowledge useful comments and conversations with Yunkyu Bang, Changyoung Kim, and Sang Jin Sin. This work was supported by the Grant from National Research Foundation of Korea under NRF-RS-2023-00246909 (THP)
and NRF-2021R1F1A1063697 (HYC).

\end{acknowledgments}



\bibliography{references.bib}{}

\section{Method}

A useful way to get a grip on the Holstein model is to integrate the phonons out and write the phonon mediated frequency dependent effective interaction between electrons as
  \begin{eqnarray}
 U_{eff} (\omega) = \frac{2g^2 \omega_0}{\omega^2 -\omega_0^2 }.
 \end{eqnarray}
In the limit $\omega_0 \rightarrow \infty $, $ U_{eff} (\omega)\rightarrow -E_b$ as discussed above.
$\gamma = \omega_0 /D $ is the adiabaticity, and one may continuously connect, by varying $\gamma$, physics of the attractive Hubbard ($\gamma \rightarrow \infty$) and Holstein models which helps understand the nature of the normal state phases and transitions between them in the Holstein and attractive Hubbard models in a unified way. We choose $\gamma =2$ in this paper which is expected to exhibit some of the attractive Hubbard model physics yet to display dynamical effects of the Holstein model.\cite{Park2019prb}
We used the Lanczos transformation for numerical calculations like Ref.\ \onlinecite{bauer_2009} with the NRG parameters $\Lambda=2.0 $, $N_S=1600$, $N_{ph}=10$, and $N=50$.

The DMFT calculations were done by considering the semi-elliptic bare DOS per spin
 \begin{eqnarray}
\rho_0 (\epsilon) = \frac{2}{\pi D^2}\sqrt{D^2-\epsilon^2}
\label{bareDOS}
 \end{eqnarray}
for the infinite dimensional Bethe lattice. Then, the self-consistency for the Weiss field ${\cal G}_{\sigma}^{-1} $ is given by
\begin{equation}\label{eq:Weiss}
  {\cal G}_{\sigma}^{-1}(\omega)= \omega +\mu - \frac{D^2}{4}G_{\sigma}^{loc}(\omega)
\end{equation}
in terms of the local Green's function $G_{\sigma}^{loc}$. It is given by the Hilbert transform of the bare DOS of Eq.\ (\ref{bareDOS}).
 \begin{eqnarray}
 G_{\sigma}^{loc}(\omega) = \frac{2}{D^2} \left( \zeta -s \sqrt{\zeta^2 -D^2} \right) ,
 \label{Glocal}
 \end{eqnarray}
where $s=sgn[Im(\zeta)]$ and $\zeta(\omega) = \omega +\mu -\Sigma(\omega)$.
The self-consistency is achieved by numerical iterations.
The renormalized DOS per spin is given by
 \begin{eqnarray}
 \rho(\omega) = -\frac1\pi Im G_\sigma^{loc} (\omega) ,
 \label{renDOS}
 \end{eqnarray}
where $Im$ stands for the imaginary part.

%

\end{document}


\setcounter{equation}{0}
\setcounter{figure}{0}
\setcounter{table}{0}
\setcounter{page}{1}
\makeatletter
\renewcommand{\theequation}{S\arabic{equation}}
\renewcommand{\thefigure}{S\arabic{figure}}
\renewcommand{\bibnumfmt}[1]{[S#1]}
\renewcommand{\citenumfont}[1]{S#1}

\title{Supplementary Information for \\ Non-Fermi liquid over extended range at zero temperature without quantum criticality }

\author{Tae-Ho Park}
\affiliation{Department of Physics and Institute for Basic Science Research, Sungkyunkwan University, Suwon 16419,
Korea}

\author{Han-Yong Choi}
\affiliation{Department of Physics and Institute for Basic Science Research, Sungkyunkwan University, Suwon 16419,
Korea}





\maketitle

\onecolumngrid

\vspace{2cm}

{\bf

S1. Comparison of total energies of coexisting solutions \\ \\

S2. Phonon spectral function $\rho_{ph} (\omega)$ and lattice fluctuations \\ \\

S3. Charge and spin susceptibility \\ \\

S4. NRG transformation energy flow \\ \\

}

\newpage

\section{S1. Comparison of total energies of coexisting solutions}\label{}

\begin{figure}[h]
\begin{center}
\includegraphics[width=0.5\linewidth]{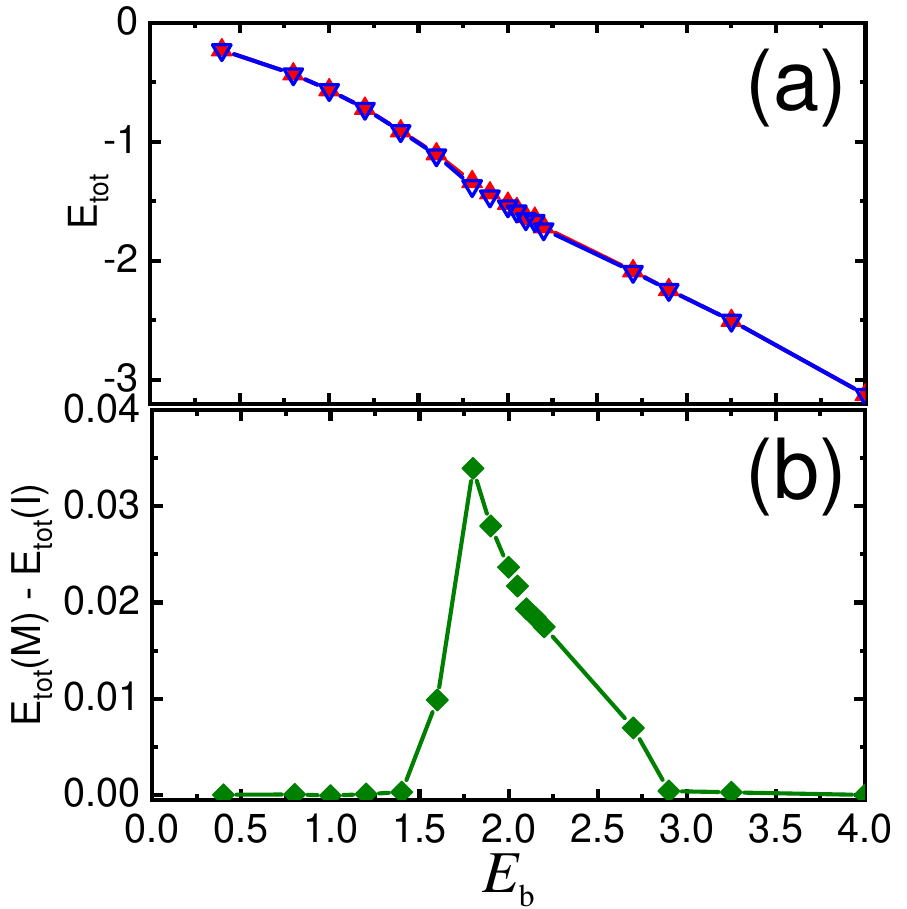}
 \caption{ The total energy of the coexisting solutions of the Hamiltonian of Eq.\ (\ref{eq:Hol}). In Fig.\ \ref{fig:energy}(a), the blue and red triangles represent the total energies of the phases from the metallic and insulating initial configurations. Their energy differences, $E_{tot}(M)- E_{tot}(I)$, are enlarged and shown in Fig.\ \ref{fig:energy}(b). The SGM is the ground state in CE1 and SGI is the ground state in CE2 region. }
\label{fig:energy}
\end{center}
\end{figure}

The DMFT calculations describe the interaction or density induced phase transitions as first order ones via coexisting regions. This has been well established since the early works on the repulsive Hubbard model.\cite{georges_1996}
For the current calculations on the Holstein model away from half-filling as well, the phase transitions are the first order transitions as presented in Fig.\ 1. The Fermi liquid metal and spin gap metal coexist in the CE1 and the spin gap metal and spin gap insulator coexist in the CE2 region in the phase diagram. A good way to obtain these two solutions is to start numerical iterations from metallic or insulating initial configurations for chosen $\mu$ and $E_b$. We calculated the total energies of the coexisting solutions to the Hamiltonian of Eq.\ (\ref{eq:Hol}) to determine the ground state. The results are shown in Fig.\ \ref{fig:energy}. The blue and red triangles in Fig.\ \ref{fig:energy}(a) represent, respectively, the results of the self-consistent calculations starting from the insulating and metallic initial configurations. The insulating (metallic) initial configurations return the SGM (FLM) in CE1 and SGI (SGM) in CE2. The energy differences between the two solutions are shown in Fig.\ \ref{fig:energy}(b). The SGM is the ground state in CE1 and SGI is the ground state in CE2 region as can be seen from the plot. The solutions from the insulating initial configurations are the ground states in all cases.


\vspace{1cm}

\section{S2. Phonon spectral function $\rho_{ph} (\omega)$ }\label{}

The phonon spectral function $\rho_{ph} (\omega)$ is the imaginary part of the Green's function,
 \ba
 \rho_{ph} (\omega) = -\frac1\pi Im D(\omega),
 \ea
and $D(\omega)$ is the phonon Green's function defined by
 \ba
 D(\omega) = \llangle a+a^\dag, a+a^\dag \rangle\!\rangle_\omega.
 \ea
Here, the correlator $\llangle ~~ \rrangle$ is defined as
 \ba
\llangle {\cal O}_1,{\cal O}_2 \rrangle_\omega &=&
\int_{-\infty}^{\infty} dt\ e^{i\omega t}\ \llangle {\cal
O}_1,{\cal O}_2 \rrangle_t ,
 \\ \nonumber
\llangle {\cal O}_1,{\cal O}_2 \rrangle_t &=& -i \theta(t) \left<
[{\cal O}_1 (t),{\cal O}_2 (0)] \right>=-i \theta(t) \left<
[{\cal O}_1 (0),{\cal O}_2 (-t)] \right>,
 \ea
where $\theta$ is the step function, $\left<~~\right>$ the
thermodynamic average, and $\left[~~\right]$ is the boson commutator.\cite{jeon_2003}

\begin{figure}[h]
\begin{center}
\includegraphics[width=0.7\linewidth]{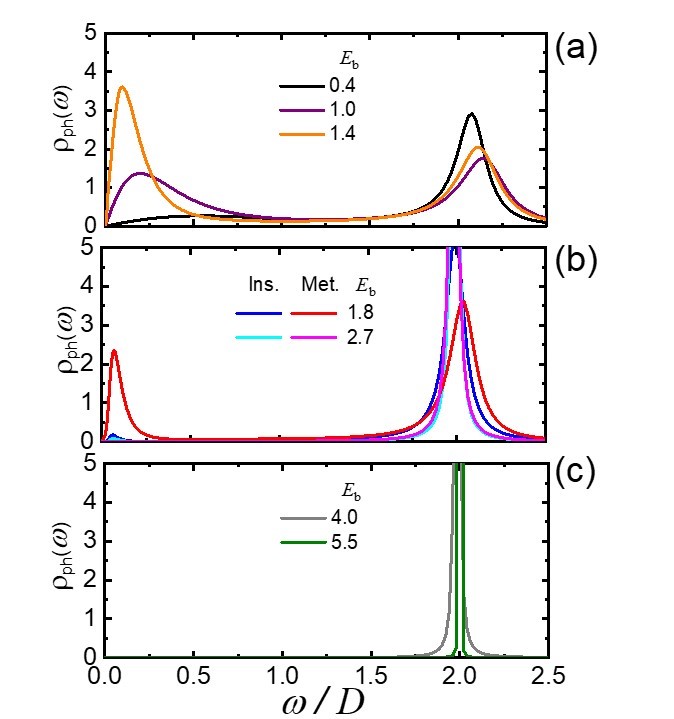}
 \caption{ The phonon spectral function $\rho_{ph} (\omega)$ as a function of the frequency for several $E_b$ along the $\mu=-0.03$ cut. Noteworthy is that a low frequency peak emerges as $E_b$ approaches the Fermi liquid-non Fermi liquid metal transition critical point $E_{c1}$. Concomitantly with the soft mode, the lattice fluctuations are greatly enhanced as shown in Fig.\ 4 in the main text. }
\label{fig:phonon}
\end{center}
\end{figure}

The results are shown in Fig.\ \ref{fig:phonon}. For a small electron phonon coupling, say for $E_b = 0.4$ in Fig.\ \ref{fig:phonon}(a), the phonon spectrum $\rho_{ph} (\omega)$ exhibits a main peak around the bare frequency $\omega_0 = 2$ and renormalized distribution of a weak intensity down to $\omega=0$. Then, as the coupling strength is increased, a low frequency peak develops with its position becoming smaller and the intensity stronger at the expense of the peak near the bare frequency. As the strength is further increased beyond $E_{c1}$, the low frequency peak is suppressed and its frequency hardens back to the bare frequency of $\omega_0 = 2$ as can be seen from Fig.\ \ref{fig:phonon}.

\begin{figure}[h]
\begin{center}
\includegraphics[width=0.5\linewidth]{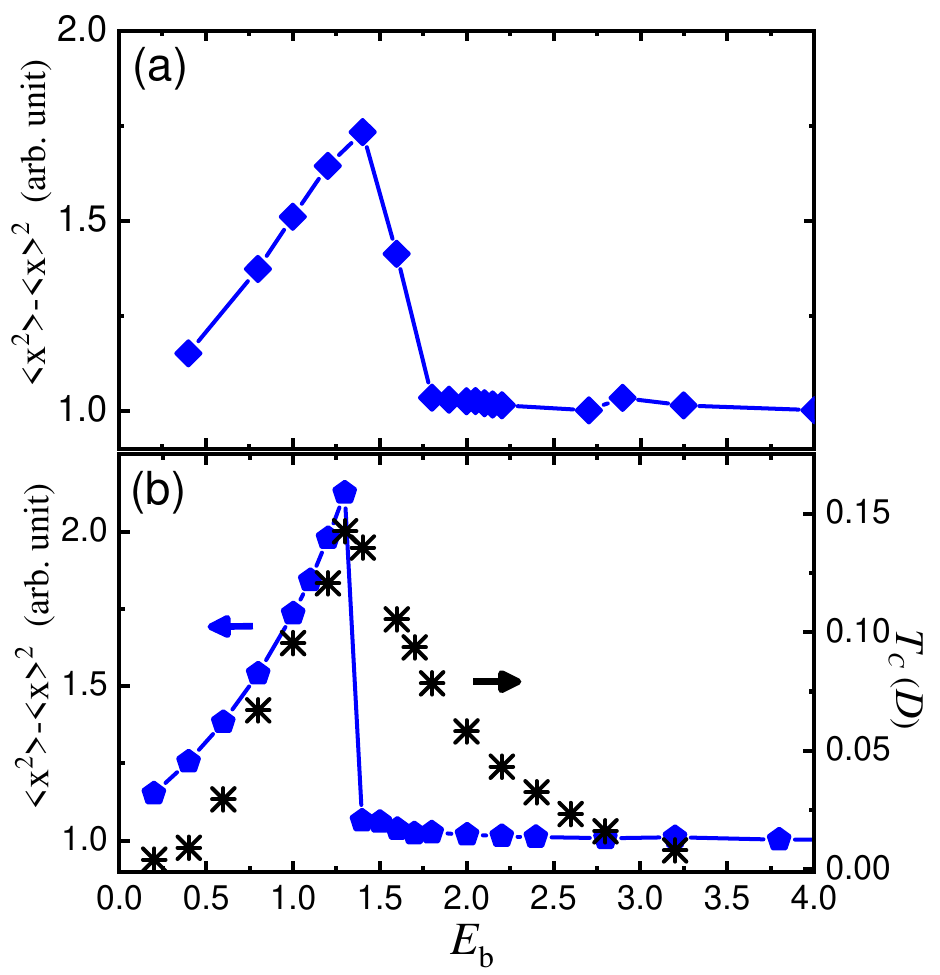}
 \caption{ (a) is along the cut $\mu = -0.03$ away from the half filling. The results for half-filling is shown in (b), and for comparison the calculated superconducting critical temperature $T_c$ is also plotted. Note that the maximum $T_c$ occurs at $E_{c1}$ in line with the attractive Hubbard model. }
\label{fig:latfluctuations}
\end{center}
\end{figure}

Accompanying this phonon softening the lattice fluctuations become enhanced near $E_{c1}$ as shown in Fig.\ \ref{fig:latfluctuations}. The lattice fluctuations are given by $\left< x^2 \right> -\left< x \right>^2 $, where $x \sim a + a^\dag $ is the displacement operator, calculated from the ground state solution. Fig.\ \ref{fig:latfluctuations}(a) is along the cut $\mu = -0.03$ away from the half filling ($n<1)$), and in (b) are shown the results for half-filling ($n=1)$) for comparison. For the latter, we also computed the superconducting critical temperature $T_c$ \cite{Park2019prb}. The maximum $T_c$ occurs at the lower critical value $E_{c1}$ as it occurs at the lower critical value $U_{c1}$ in the attractive Hubbard model \cite{toschi_2005,toschi_2005b}.

\vspace{2cm}

\section{S3. Spin and Charge susceptibility }\label{}

Here, we show that the  spin gap from the spin susceptibility spectra is exhibited in both the spin gap metal (SGM) and the spin gap insulating (SGI) states, while the charge gap from the charge susceptibility spectra appears  only in the insulating state. The charge gap is consistent with the single particle gap as shown in Fig.\ 2 in the main text.
The spin and charge susceptibilities are calculated from the spin-spin and  density-density correlations respectively. Here, we consider the longitudinal spin-spin correlation function, $\chi_s(\omega)=\llangle \hat{S}_z, \hat{S}_z \rangle\!\rangle_\omega$, where $\hat{S}_z$ is a longitudinal component of the spin operator  as $\hat{S}_z=\frac{1}{2}(\hat{n}_{\uparrow}-\hat{n}_{\downarrow})$ and the density-density correlation function is $\chi_c(\omega)=\llangle \hat{n}, \hat{n} \rangle\!\rangle_\omega$, where $\hat{n}=\hat{n}_{\uparrow}+\hat{n}_{\downarrow}$ and $\hat{n}_{\sigma}$ is the electron density operator for spin $\sigma$.

In the spectral form, the imaginary part of the spin-spin and the density-density correlation functions are calculated by the Lehmann representation as following,
\ba \label{eq:ImSpin}
-\frac{1}{\pi} Im \chi_s(\omega) &=& \frac {1}{Z} \sum_{n m} \lvert  \langle  n \lvert  \hat{S}_z  \rvert m \rangle \rvert^2 \delta ( \omega - (E_n -E_m))(e^{-\beta E_n} - e^{-\beta E_m}),
\\ \label{eq:ImCharge}
-\frac{1}{\pi} Im \chi_c(\omega) &=& \frac {1}{Z} \sum_{n m} \lvert  \langle  n \lvert  \hat{n} \rvert m \rangle \rvert^2 \delta ( \omega - (E_n -E_m))(e^{-\beta E_n} - e^{-\beta E_m}),
\ea
where, $Z$ is the partition function and $\beta= 1/k_B T$ ($k_B$ is the Boltzmann constant and $T$ is the temperature). $E_n$ and $E_m$ are eigenenergies obtained from the numerical renormalization group (NRG) calculation and  $\lvert n \rangle = \lvert Q, S_{z}, w_n \rangle $ and $\lvert m \rangle = \lvert Q, S_{z}, w_m \rangle $ represent the corresponding eigenstates, which are expressed by quantum numbers, the charge $Q$ and the projected spin $S_z$ and $w_n$ ($w_m$) is the labeling number for the diagonalized states on a $Q$ and $S_z$ block matrix.\cite{Hofstetter_2000}
These eigenenergies and eigenstates are obtained from the recursive form of the Hamiltonian representing the fictitious chain in the NRG calculation as following,
\begin{equation}\label{eq:NRGrecur}
{\cal H}_{N}=\Lambda^{1/2}{\cal H}_{N-1}+\xi_{N-1} \sum_{\sigma} \left( f^{\dag}_{(N-1) \sigma}f_{N\sigma}+ f^{\dag}_{N \sigma}f_{(N-1)\sigma} \right)+\epsilon_N \sum_{\sigma} f^{\dag}_{N \sigma} f_{N\sigma} ,
\end{equation}
where $\Lambda$ is the logarithmic scaling parameter for the NRG calculation and $f_{N \sigma}$ ($f^{\dagger}_{N \sigma}$) is the annihilation (creation) operator of the particle on the $N$'th site with a spin $\sigma$.  $\xi_{N-1}$ is the tight-binding hopping parameter with the logarithmic scale for the chain and $\epsilon_N$ is the on-site energy term given by the particle-hole asymmetry, and both of them are obtained through the combination with the dynamical mean field theory (DMFT) self-consistent calculation. As we diagonalize the Hamiltonian Eq.\ \ref{eq:NRGrecur}, the eigenenergies $E_{N}(Q,S_z,w)$ and eigenstates $\lvert Q, S_{z}, w \rangle = \sum_{i,r} U_{Q,S_z}(w;ri)\lvert Q, S_{z}, r,i \rangle$ are obtained, where $r$ is the state labeling number before the diagonalization and $i$ indicates four spin configuration on a site as empty, single occupancy and doubly occupied states (i.e. $1: 0$, $2: \uparrow$, $3:\downarrow$, $4:\uparrow \downarrow$).
Then, the matrix elements in Eq.\ \ref{eq:ImSpin} and \ref{eq:ImCharge} can be calculated as
\ba  \label{eq:MatrixElSz}
\left< Q, S_{z}, w_n \lvert  \hat{S}_z \rvert Q, S_{z}, w_m\right> &=& \sum_{r,r',i} U_{Q, S_{z}}\left(w_n,;r, i \right) U_{Q, S_{z}}\left(w_m,;r', i \right) \left< Q, S_{z}, r,i \lvert  \hat{S}_z \rvert Q, S_{z}, r',i\right>,
\\  \label{eq:MatrixElQ}
\left< Q, S_{z}, w_n \lvert  \hat{n} \rvert Q, S_{z}, w_m\right> &=& \sum_{r,r',i} U_{Q, S_{z}}\left(w_n,;r, i \right) U_{Q, S_{z}}\left(w_m,;r', i \right) \left< Q, S_{z}, r,i \lvert  \hat{n} \rvert Q, S_{z}, r',i\right>,
\ea
and the matrix elements of $\hat{S}_z$  in the righthand side in Eq.\ \ref{eq:MatrixElSz} are given by the recursion relations,
\ba
\left< Q, S_{z}, r,1\lvert  \hat{S}_z \rvert Q, S_{z}, r',1\right>_{N} &=& \left< Q+1, S_{z}, r\lvert  \hat{S}_z \rvert Q+1, S_{z}, r'\right>_{N-1},
\\ \nonumber
\left< Q, S_{z}, r,2\lvert  \hat{S}_z \rvert Q, S_{z}, r',2\right>_{N} &=& \left< Q, S_{z}-\frac{1}{2}, r\lvert  \hat{S}_z \rvert Q, S_{z}-\frac{1}{2}, r'\right>_{N-1},
\\ \nonumber
\left< Q, S_{z}, r,3\lvert  \hat{S}_z \rvert Q, S_{z}, r',3\right>_{N} &=& \left< Q, S_{z}+\frac{1}{2}, r\lvert  \hat{S}_z \rvert Q, S_{z}+\frac{1}{2}, r'\right>_{N-1},
\\ \nonumber
\left< Q, S_{z}, r,4\lvert  \hat{S}_z \rvert Q, S_{z}, r',4\right>_{N} &=& \left< Q-1, S_{z}, r\lvert  \hat{S}_z \rvert Q-1, S_{z}, r'\right>_{N-1},
\ea
and the matrix elements of $\hat{n}$ in the righthand side in Eq.\ \ref{eq:MatrixElQ} are given by the recursion relations,
\ba
\left< Q, S_{z}, r,1\lvert  \hat{n} \rvert Q, S_{z}, r',1\right>_{N} &=& \left< Q+1, S_{z}, r\lvert  \hat{n} \rvert Q+1, S_{z}, r'\right>_{N-1},
\\ \nonumber
\left< Q, S_{z}, r,2\lvert  \hat{n} \rvert Q, S_{z}, r',2\right>_{N} &=& \left< Q, S_{z}-\frac{1}{2}, r\lvert  \hat{n} \rvert Q, S_{z}-\frac{1}{2}, r'\right>_{N-1},
\\ \nonumber
\left< Q, S_{z}, r,3\lvert  \hat{n} \rvert Q, S_{z}, r',3\right>_{N} &=& \left< Q, S_{z}+\frac{1}{2}, r\lvert  \hat{n} \rvert Q, S_{z}+\frac{1}{2}, r'\right>_{N-1},
\\ \nonumber
\left< Q, S_{z}, r,4\lvert  \hat{n} \rvert Q, S_{z}, r',4\right>_{N} &=& \left< Q-1, S_{z}, r\lvert  \hat{n} \rvert Q-1, S_{z}, r'\right>_{N-1},
\ea
with initial conditions $ \left< 0, \pm \frac{1}{2} \lvert \hat{S}_z \rvert 0, \pm \frac{1}{2} \right> = \pm 1/2$ for $\hat{S}_z$ and
$ \left< 0, \pm \frac{1}{2} \lvert \hat{n} \rvert 0, \pm \frac{1}{2} \right> = 1$ and $ \left< 1, 0 \lvert \hat{n} \rvert 1, 0 \right> = 2$ for $\hat{n}$.

The real part of the correlation function is obtained from the Kramers-Kronig relation as
\begin{equation}\label{eq:KK}
Re \chi(\omega) = P \int_{-\infty}^{\infty} \frac{d \omega '}{\pi} \frac{Im \chi (\omega ')}{\omega ' - \omega}.
\end{equation}
We calculated  the spectra of the spin susceptibility $\chi_s(\omega)$ and the charge susceptibility $ \chi_c(\omega)$. The charge and spin spectra of the ground state phases along the $\mu=-0.03$ cut are shown in Fig.\ \ref{fig:spingap}(a) and (b) in the main text. $E_b = 1.9$ represents SGM of the CE1 region, $E_b = 1.0$ \& 1.4 refer to the FLM, and $E_b = 2.7$ \& 3.25 refer to the SGI phases.

\vspace{2cm}


\section{S4. NRG transformation energy flow  }\label{NRG_flow}

\begin{figure}[h]
\begin{center}
\includegraphics[width=\linewidth]{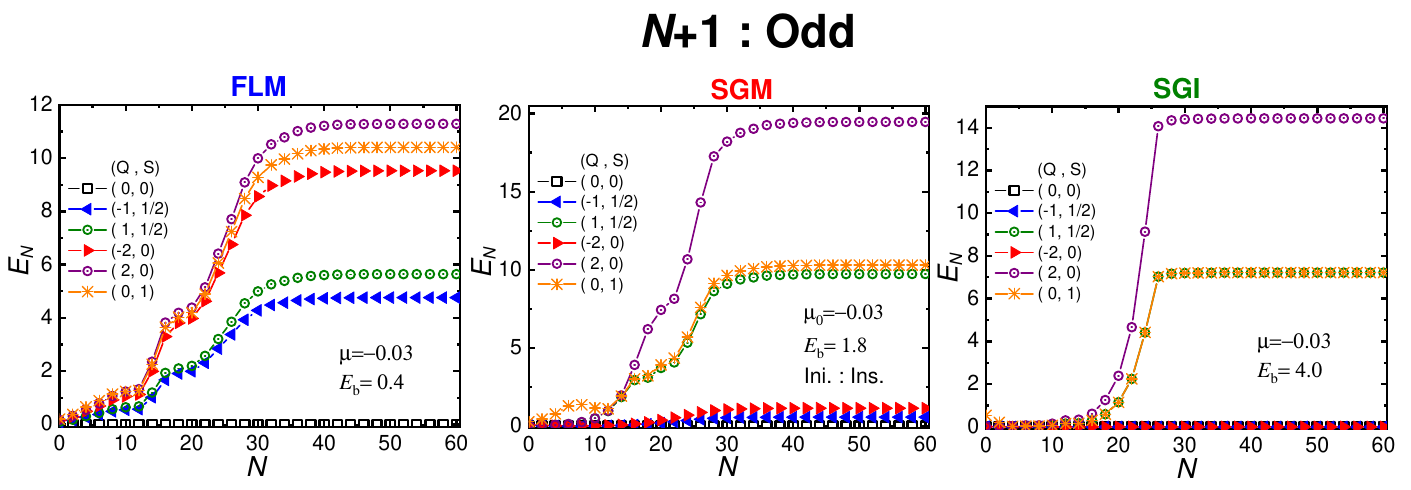}
 \caption{ The NRG energy flow of eigenstates under $T^2$ labeled by the charge and spin quantum numbers, $Q$ and $S$, for odd $N+1$. The NRG transformation $T$ does not have a fixed point, but $T^2$ does. The distinct difference of ground state degeneracy of the three fixed points guarantee that there exist three phases in the lattice problem. }
\label{fig:NRGflow}
\end{center}
\end{figure}

\bibliography{references.bib}{}